\documentclass[letterpaper, 10 pt, conference]{ieeeconf}
\IEEEoverridecommandlockouts  
\usepackage{amsmath}
\usepackage{amssymb}
\usepackage{amsthm}
\usepackage{mathtools}
\usepackage{array}
\usepackage{graphicx}
\usepackage{textcomp}
\usepackage{stfloats}
\usepackage{subfigure}
\usepackage{color}
\usepackage{url}
\usepackage{verbatim}
\usepackage{tikz}
\usepackage{xcolor}
\usepackage{mdframed}
\usepackage{hyperref}
\usepackage{balance}
\usepackage{cite}

\usepackage[linesnumbered,ruled,vlined]{algorithm2e}

\usepackage{courier}

\newcommand{\qi}[1]{{\color{black} #1}}
\newcommand{\tang}[1]{{\color{black} #1}}

\newtheorem{definition}{Definition}

\newtheorem{lem}{Lemma}

\newcommand{\Iff}{\mathrel{\text{iff}}}

\title{Integrating Opinion Dynamics into Safety Control for \\ Decentralized Airplane Encounter Resolution}
\author{Shuhao Qi$^{1}$ , Zhiqi Tang$^2$ , Zhiyong Sun$^{3}$ and Sofie Haesaert$^{1}$ \vspace{-1em}
\thanks{This work is supported by the European project SymAware under grant No. 101070802, the European project COVER under grant No. 101086228, and the Dutch NWO Veni project CODEC under grant No. 18244. Corresponding author: \textit{Zhiyong Sun}.}
\thanks{$^{1}$ S. Qi and S. Haesaert are with the Department of Electrical Engineering, Eindhoven University of Technology, Eindhoven, The Netherlands.
        {\tt\small \{s.qi, s.haesaert\}@tue.nl}}
\thanks{$^{2}$Z. Tang is with the Division of Decision and Control Systems, KTH Royal Institute of Technology, Stockholm, Sweden. {\tt\small \{ztang2@kth.se\}}}
\thanks{$^{3}$Z. Sun is with the College of Engineering, Peking University, Beijing, China. {\tt\small \{zhiyong.sun@pku.edu.cn\}}}
}

\begin{document}
\maketitle

\begin{abstract} 
As the airspace becomes increasingly congested, decentralized conflict resolution methods for airplane encounters have become essential. While decentralized safety controllers can prevent dangerous midair collisions, they do not always ensure prompt conflict resolution. As a result, airplane progress may be blocked for extended periods in certain situations. To address this blocking phenomenon, this paper proposes integrating bio-inspired nonlinear opinion dynamics into the airplane safety control framework, thereby guaranteeing both safety and blocking-free resolution. In particular, opinion dynamics enable the safety controller to achieve collaborative decision-making for blocking resolution and facilitate rapid, safe coordination without relying on communication or preset rules. Extensive simulation results validate the improved flight efficiency and safety guarantees.  This study provides practical insights into the design of autonomous controllers for airplanes.

\end{abstract}
\section{Introduction}

Rising demand for aviation is crowding airspace, leading to more frequent airplane encounters~\cite{aam2022}.  Since centralized air traffic management scales poorly to large-scale traffic~\cite{de2021decentralized}, encounter conflicts must be resolved autonomously through decentralized coordination among the airplanes involved. Although many decentralized approaches depend on communication and fixed rules, they are not always reliable because inter-airplane communication is vulnerable to disturbances, delays, and language ambiguity, and preset rules might be incompatible among different airplanes~\cite{master2020}. Therefore, a decentralized conflict resolution method should be robust to communications failures and incompatible rules, serving as a fallback in emergent situations. However, developing such a decentralized resolution approach that guarantees both safety and successful resolution remains an open challenge.

\begin{figure}[t]
    \centering
    \includegraphics[width=0.45\textwidth]{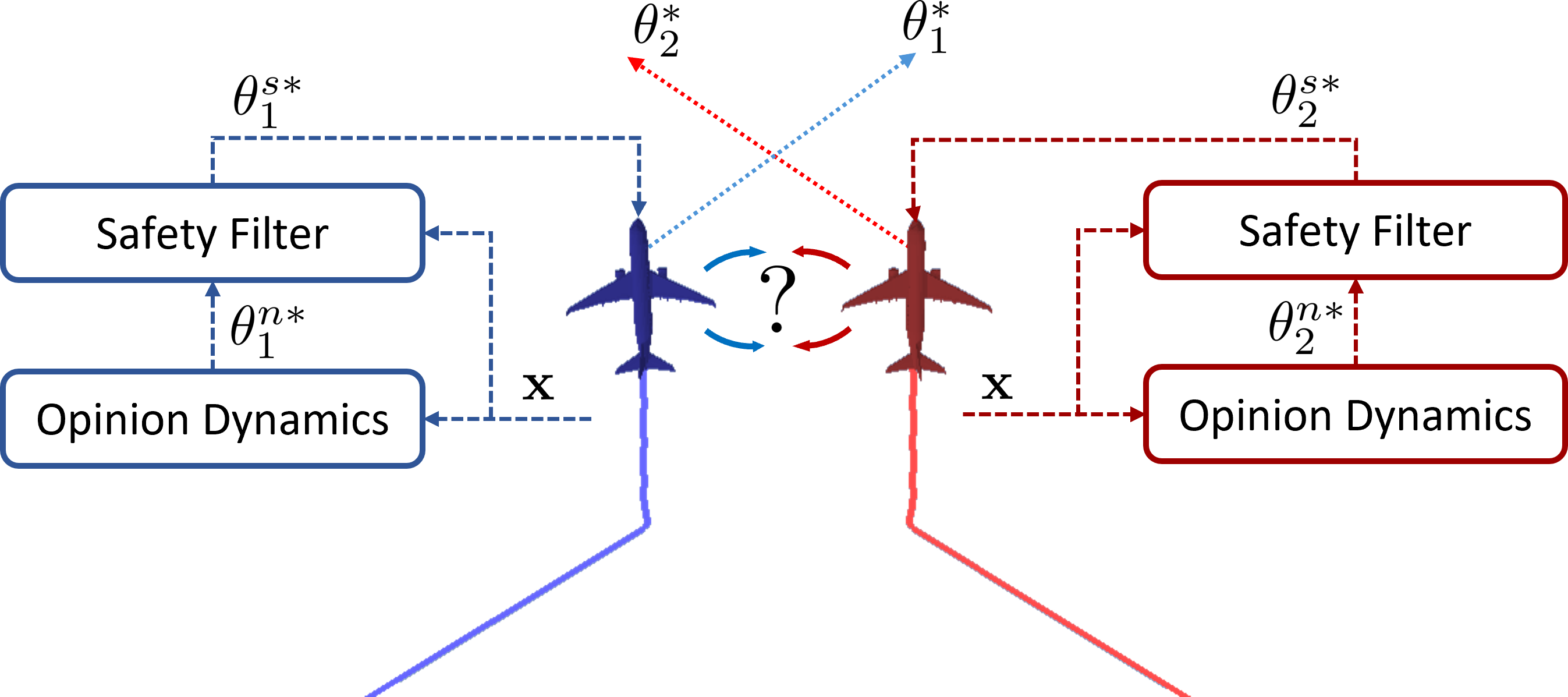}
    \caption{Control framework for two-airplane encounters with safety guarantee and blocking-free resolution. \qi{$\theta^*_i$ is the desired cruising angle (which may be unsafe and causes blocking), $\theta^{n*}_i$ is the opinion-guided heading angle designed to facilitate cooperation and resolve blocking, and $\theta^{s*}_i$ is the safe heading angle adjusted by the safety filter to ensure safety. }} 
    \label{fig:overview}
\end{figure}

Over the past decade, safety control theory has developed through advances in control barrier function (CBF)~\cite {wang2017safety}, safe-reachable set~\cite{fisac2018general}, etc. These advances facilitated the design of decentralized controllers with formal safety guarantees for multi-robot systems~\cite{wang2017safety, chen2023multi, weng2022convergence}. However, safety and task completion (control-theoretic stability) may conflict in symmetric situations, leading to undesired equilibria~\cite{reis2020control, grover2020does}. When this occurs, task completion is sacrificed to ensure safety, resulting in pathological phenomena like deadlocks~\cite{grover2023before}, in which all robots stop indefinitely before reaching targets. Although fixed-wing airplanes cannot stop in place like ground robots,  recent air-traffic simulations have revealed a deadlock-like behavior~\cite{nlr24report} when every airplane is equipped with detect and avoid (DAA) systems~\cite{monk2019detect} to ensure safety. In this phenomenon, two airplanes following instructions of DAA systems repeatedly strive but fail to bypass each other, resulting in parallel flight for a finite time, as illustrated by the trajectories in Fig.~\ref{fig:overview}. Like deadlock, this phenomenon stems from the conflict between task completion and safety. Our previous work~\cite{qi2024avoiding} formally characterized this new phenomenon and termed it as \textit{blocking}.  Blocking not only delays progress but also increases the risk of near-miss collisions between airplanes, particularly under realistic uncertainty. Because DAA systems are widely deployed to increase the autonomy of airplanes, eliminating such blocking behaviors is imperative. Resolving both deadlock and blocking phenomena requires coordination among the participating agents. Moreover, such cooperative decisions must be reached swiftly, as indecision can be catastrophic for high-speed airplanes. However, achieving swift and safe coordination without communication is challenging, even for a two-airplane encounter.

Drawing on insights from social and biological sciences, the nonlinear opinion dynamics (NOD) framework has recently been adopted to model interactions among multiple agents~\cite{bizyaeva2022nonlinear, leonard2024fast}. In NOD, each agent’s opinion state numerically represents its degree of agreement with a decision. NOD’s inherent bifurcation property~\cite{leonard2024fast} makes each agent sensitive to slight changes of others’ states, enabling the group to make collaborative decisions rapidly. With a well-designed NOD, the agents can stabilize toward the desired collaborative decision as the interaction evolves. Thus, NOD shows promise of fast deadlock resolution~\cite{cathcart2023proactive}. 

Motivated by these findings, this paper tailors NOD for airplanes and leverages it to resolve pathological blocking phenomena. Unlike earlier work that applied NOD to robot navigation~\cite {cathcart2023proactive, cathcart2024spiking}, we employ NOD as a complement to eliminate blocking within a safety control framework, as shown in Fig.~\ref{fig:overview}. The resulting scheme lets two airplanes resolve conflicts safely and collaboratively without fixed rules or direct communication. Analytical proofs and extensive simulations confirm safety and blocking-free resolution in pairwise encounters. The proposed method addresses the shortcomings of DAA systems identified in~\cite{nlr24report}, providing practical insights for their further advancement.



In this paper, vector variables will be given as bold symbols,  $\mathbf{x}\!\in\! \mathbb R^n$, while scalars will be denoted as $x\!\in\!\mathbb R$. $\|\cdot\|$ denotes Euclidean norm of a vector. Furthermore, we define the following angular normalization operator, which maps an angle $a \!\in\!\mathbb R$ to the range  $[-\pi, \pi)$,
\begin{equation}
    \measuredangle(a) = (a + \pi) \% 2 \pi - \pi,
    \label{eq:operator}
\end{equation}
\noindent where $\%$ is the modulo operator.

\section{Preliminary and Problem Formulation}
\subsection{Two-airplane system}
Consider an air traffic scenario involving $2$ fixed-wing airplanes, denoted by $\mathcal{A}\!=\!\{A_1, A_2\}$. The horizontal behavior of an airplane $A_i$ is characterized by its position $\mathbf{p}_i(t) = [p_i^x(t), p_i^y(t)]^T\!\in\!\mathbb{R}^2$ and its heading angle $\theta_i(t)$, with respective shorthand notation $\mathbf{p}_i$ and $\theta_i$. Define $\mathbf x_i\!=\![\mathbf p_i^\top, \theta_i]^\top$ and a simplified airplane model is adopted here using unicycle dynamics with a constant forward velocity $v\!>\!0$~\cite{sun2021collaborative},
\begin{equation}
\dot{\mathbf{x}}_i = \begin{bmatrix} \dot{p}_i^x \\ \dot{p}_i^y \\ \dot{\theta}_i \end{bmatrix} = \begin{bmatrix}
    v \cos (\theta_i)\\
    v \sin (\theta_i)\\
    a_i
\end{bmatrix},\quad i\in\{1,2\}.
\label{eq:sys}
\end{equation}

\noindent The stacked state of the two airplanes is denoted by $\mathbf{x}\!=\![\mathbf{x}_1^\top, \mathbf{x}_2^\top]^\top$. Each airplane aims to track its desired cruising angle denoted by $\theta_i^{*}$. 

\begin{figure}[t]
    \centering
    \includegraphics[width=0.25\textwidth]{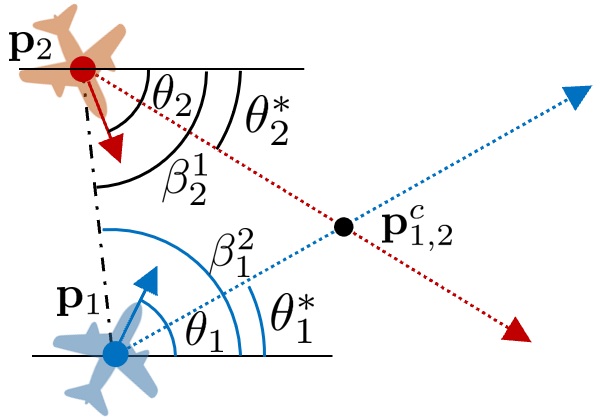}
    \caption{Geometric relationships in the two-airplane encounter.} 
    \label{fig:sketch}
\end{figure}

There are situations where two airplanes might encounter each other, with the risk of collision. This is referred to as an encounter situation. A necessary but insufficient condition for encounter scenarios is the existence of intersection points between the cruising paths of two airplanes, like $\mathbf{p}^c_{1,2}$ in Fig.~\ref{fig:sketch}. Hence, to avoid collisions, the airplanes must maintain a safe margin $r\!>\!0$. That is, $\forall t \!\geq\! 0$,
\begin{equation}
\|\mathbf{p}_1(t)-\mathbf{p}_2(t)\| \geq r.\label{eq:safetyreq}
\end{equation}

\qi{In this work, our focus is to design a safe heading angle $\theta_i^{s*}$  to ensure safety without compromising the cruising to the desired angle, $\theta_i^*$. We use a high-gain tracking controller 
\begin{equation}\label{ea:ai}
a_i=-k(\theta_i-\theta_i^{s*})+\dot \theta_i^{s*},
\end{equation}
where $k$ is a sufficiently large positive gain such that $\theta_i$ approaches $\theta_i^{s*}$ instantaneously (i.e., $\theta_i\approx \theta_i^{s*}$). Under this setting, the model~\eqref{eq:sys} simplifies to  
\begin{equation} \label{eq:simplied-sys}
    \dot{\mathbf{p}}_i\!=\!\begin{bmatrix} v\text{cos}(\theta_i^{s*})\\ v\text{sin}(\theta_i^{s*}) 
\end{bmatrix},\quad i\in \{1,2\}.
\end{equation} }

\subsection{CBF-based safety filter}

Realistic DAA systems typically rely on complex methodologies, such as dynamic programming and extensive lookup tables~\cite{nlr24report}, which pose significant challenges for formal analysis. Therefore, we instead employ an analytically tractable safety filter~\cite{CSM23} for the model in~\eqref{eq:simplied-sys}, which emulates the functionality of DAA systems to ensure safety. From a control-theoretic perspective, the safety property is formally captured by forward invariance~\cite{wang2017safety}. CBF~\cite{ames2019control} is a convenient approach for enforcing forward invariance in a dynamical system and is defined as follows.
\begin{definition}[Control barrier function, CBF]
Let $\mathcal{C} \subset \mathbb{R}^n$ be a super level set of a continuously differentiable function $h: \mathbb{R}^n \!\rightarrow\!\mathbb{R}$. Then $h$ is a valid CBF that ensures the forward invariance of $\mathcal{C}$ for a nonlinear system $\dot{x}\!=\!f(x,u)$ if there exists a continuous function $\alpha(\cdot):\mathbb{R} \!\rightarrow \!\mathbb{R}$ such that for all $x \!\in\!\mathbb{R}^n$, there exists a
control input $u\!\in\!\mathbb{R}^m$ satisfying: 
$$
\textstyle \alpha(h(x)) + \frac{d }{d x}[h(x)] f(x,u)\geq 0. 
$$ 
where $\alpha(\cdot)$ is strictly monotonically increasing with $\alpha(0)\!=\!0$. 
\end{definition} 
\noindent In this paper, we define a CBF as 
$h(\mathbf{p}_1, \mathbf{p}_2)\!:=\!\|\mathbf{p}_1-\mathbf{p}_2\|^2 - r^2 \geq 0$
and, for simplicity, we use a linear function $\alpha(x)\! = \! \alpha x, \alpha\!\in\!\mathbb R^+$. Referring to~\cite{wang2017safety, grover2023before}, we derive the decentralized CBF condition for airplane $A_i$ as follows,
\begin{equation}
    g(\mathbf{p}_i, \mathbf{p}_j, \theta_i) := \frac{\alpha}{2} h(\mathbf{p}_1, \mathbf{p}_2)\! +\! 2v(\mathbf{p}_i\!-\!\mathbf{p}_j)^T \!\begin{bsmallmatrix} \text{cos}(\theta_i)\\ \text{sin}(\theta_i)
\end{bsmallmatrix} \geq 0,
    \label{eq:cbf}
\end{equation}
where $i, j\in\{1,2\}$ and $i\ne j$. Such CBF conditions can ensure safety in a decentralized manner based on the half-responsibility assignment. With this constraint, a safety filter, denoted by $\mathcal{F}(\mathbf{x}, \theta_i^*)$, is formulated as a quadratic programming problem, 
\begin{align} 
 \textstyle  \mathcal{F}(\mathbf{x}, \theta_i^*)   :=&\arg\underset{\qi{\theta_i^{s*}} \in \mathbb{R}}{\operatorname{min}} && \frac{1}{2}\|\qi{\theta_i^{s*}} -\theta_i^* \|^2  \label{eq:sf}\\
&\, \text{s.t.} &&  g(\mathbf{p}_i, \mathbf{p}_j, \qi{\theta_i^{s*}} ) \geq 0, \, j=\{1,2\}/i. \notag
\end{align}
Like a DAA system, this safety filter can detect unsafe cruising control inputs and adjust them with minimal interference. Its feasible solutions are provably safe.



\subsection{Blocking phenomenon}
\begin{figure}[t]
    \centering
    \includegraphics[trim=40 0 30 0, clip, width=0.42\textwidth]{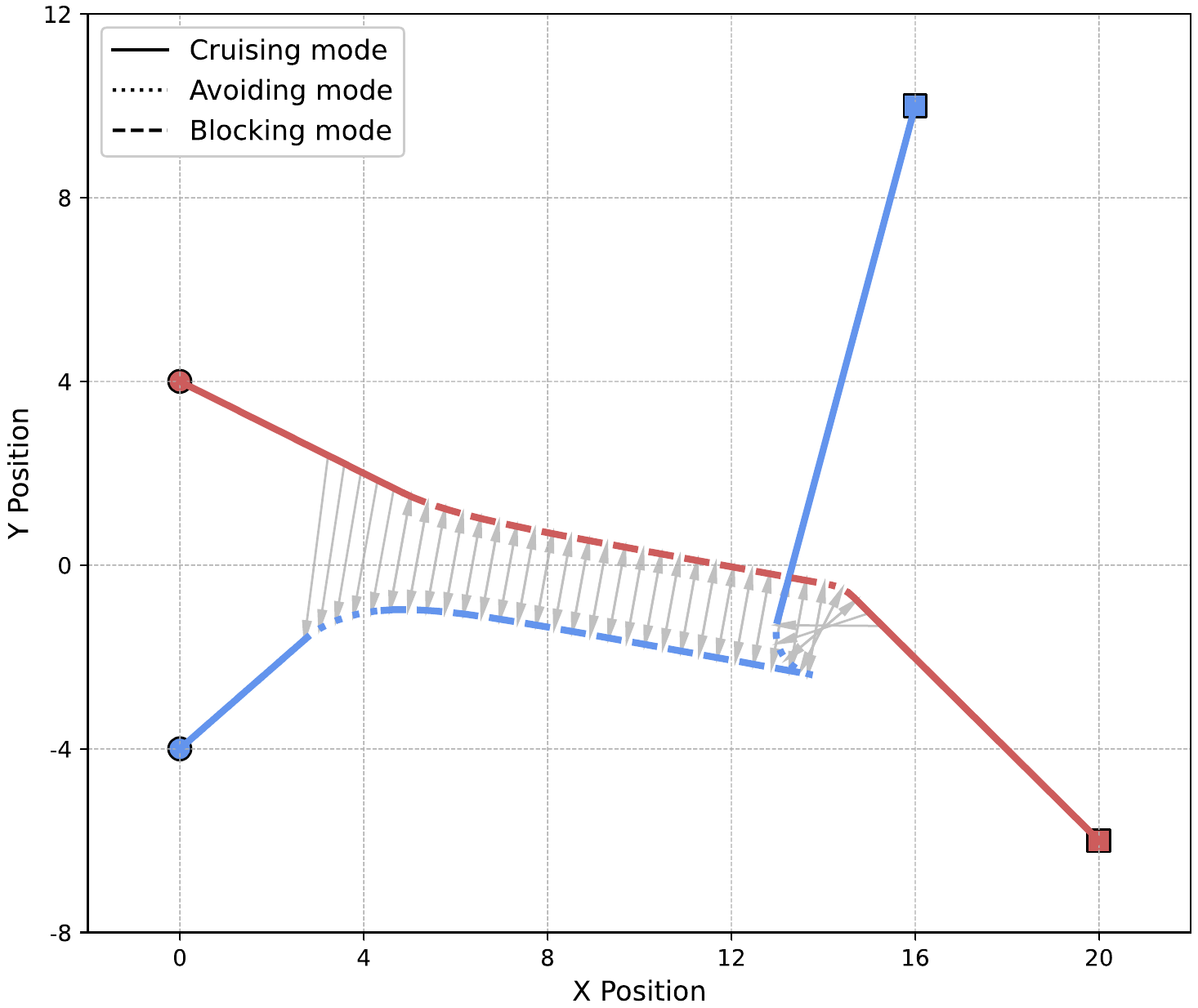} 
    \caption{Trajectories of two airplanes switching among three flight modes. Gray arrows indicate the activation of safety filters. Circle and square markers denote initial and target positions, respectively. } 
  \label{fig:blocking_sim} 
\end{figure}

In an encounter situation, the safety filters activate as airplanes approach the encounter point, $\mathbf{p}^c_{1,2}$. In certain situations, the safety filter may compromise the task completion (i.e., convergence of $\theta_i$ to the desired angle $\theta^*_i$) to ensure safety.  As mentioned earlier, a pathological phenomenon named as~\textit{blocking} has recently been observed in air traffic simulations~\cite{nlr24report}, as shown in Fig.~\ref{fig:blocking_sim}. This phenomenon has been thoroughly characterized and analyzed in~\cite{qi2024avoiding}. 
Define $\beta_i^j\in \mathbb R$ as the bearing angle from $A_i$ to $A_j$ (see Fig.~\ref {fig:sketch}), such that $\begin{bsmallmatrix} \text{cos}(\beta_i^j)\\ \text{sin}(\beta_i^j) \end{bsmallmatrix}\!=\!\frac{\mathbf{p}_j - \mathbf{p}_i}{\|\mathbf{p}_j - \mathbf{p}_i\|}$, with $i,j\in\{1,2\}$ and $j\ne i$. The \textit{blocking mode} is formally defined as follows. 
\begin{definition}[Blocking mode]\label{def:blocking}
Considering a two-airplane system $\mathcal{A}\!=\!\{A_1, A_2\}$, $A_i$ is in blocking mode at a time instant $T$ if its safety filter is active, i.e.,
 $\mathcal{F}(\mathbf{x}(T), \theta_i^*)\!\neq\! \theta_i^*$, and the relative bearing angle is constant, i.e.,
$\dot{\beta}_i^j\!(T)=\!0$.  
\end{definition}

\noindent The derivative of $\beta_i^j$ can be calculated by $\dot{\beta}_i^j\!:=\!\frac{ (\mathbf{p}_j - \mathbf{p}_i) \times (\mathbf{\dot{p}}_j - \mathbf{\dot{p}}_i)}{\|\mathbf{p}_j - \mathbf{p}_i\|^2}$, which is a scalar, due to the cross product in two-dimensional space. In addition,  $A_i$ is in the \textit{cruising mode} if $\mathcal{F}(\mathbf{x}, \theta_i^*)\!=\!\theta_i^*$, or in \textit{avoiding mode} if $\mathcal{F}(\mathbf{x}, \theta_i^*)\!\neq\!\theta_i^*$ and $\dot{\beta}_i^j\!\neq\!0$. As shown in Fig.~\ref{fig:blocking_sim}, an airplane switches among the three modes throughout the encounter.

\smallskip

This paper aims to develop a decentralized resolution method for the blocking mode defined in Def~\ref{def:blocking} in two-airplane encounters without relying on communications or preset rules. In Sec.~\ref {sec:2airplane}, we first design an opinion dynamics model for airplane encounters, and integrate it within the safety filter framework~\eqref{eq:sf}. We then provide a theoretical analysis of the opinion-guided approach for resolving blocking phenomena in Sec.~\ref{sec:analysis}, followed by extensive experiments in Sec.~\ref{sec:exper} to validate its safety and blocking-free resolution in two-airplane encounter scenarios.

\section{Blocking Resolution using Nonlinear Opinion Dynamics}\label{sec:2airplane}
In this section, we analyze the blocking conditions during airplane encounters, and then, inspired by~\cite{cathcart2023proactive, leonard2024fast}, we propose a nonlinear opinion dynamics mechanism to break blocking while guaranteeing safety.

\subsection{Blocking condition}

\begin{figure}[tb]
    \centering
    \includegraphics[width=0.37\textwidth]{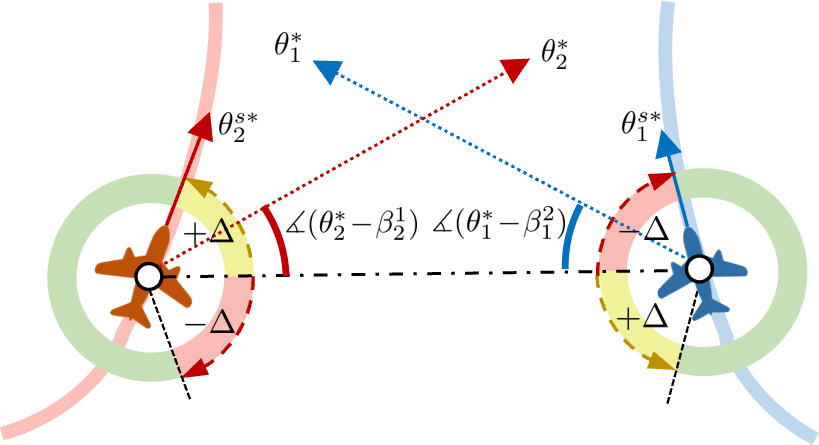}
    \caption{Explicit solution of safety filters and blocking condition. }
    \label{fig:solution}
\end{figure}

In an encounter scenario involving two airplanes $\mathcal{A}\!=\!\{A_1, A_2\}$, when the airplanes become closer, their safety filter~\eqref{eq:sf} will activate to adjust desired cruising angle $\theta^*_i$ to safe cruising angle  $\theta^{s*}_i$. Fig.~\ref{fig:solution} illustrates how the safety filter functions. For the desired angles $\theta^*_i$,  those that lie within the green arc are considered safe, whereas those within the red or yellow arcs are classified as unsafe. The red and yellow arcs represent the negative and the positive corrections, denoted by $+\Delta$ and $-\Delta$, respectively, indicating the intentions to bypass the other airplane on the right and left. They are symmetric about the axis aligned with the vector $\boldsymbol p_1-\boldsymbol p_2$. Hence, the unsafe set is characterized by $\|\measuredangle(\theta^*_i -\beta_i^j)\|\!<\!\Delta$, for $i,j\!\in\!\{1,2\}$ and $i\!\neq\!j$. When the airplanes are far apart, $\Delta\!=\!0$. As the airplanes approach each other, $\Delta$ gradually converges to $\frac{\pi}{2}$. Due to the minimal interference manner, the safety filter will correct the unsafe cruising angle $\theta_i^*$ to the nearest safe angle, i.e., $\beta_i^j\! +\!\Delta$ or $\beta_i^j\!- \!\Delta$.  The function of safety filters $\theta_i^{s*}\!=\!\mathcal{F}(\mathbf{x}, \theta_i^{*})$ in Eq.\eqref{eq:sf} is formally captured by the following explicit solution
\begin{equation}
\begin{aligned}
      \theta_i^{s*} &= 
    \begin{cases}
      \beta_i^j \pm \Delta, & \theta_i^* = \beta_i^j \\
    \beta_i^j - \Delta, & \measuredangle(\theta_i^* - \beta_i^j) \in (-\Delta, 0)  \\
    \beta_i^j + \Delta, & \measuredangle(\theta_i^* - \beta_i^j) \in (0, \Delta)  \\
    \theta_i^*, & otherwise\\
    \end{cases}
\end{aligned}
\label{eq:sol}
\end{equation} 
where $i,j\!\in\!\{1,2\}$ and $i\!\neq\!j$, $\measuredangle(\cdot)$ operator is defined in Eq.~\eqref{eq:operator}, and $\Delta\!=\! \text{cos}^{-1} \left(\text{min}(1, \frac{\alpha h(\mathbf{p}_i, \mathbf{p}_j)}{4v \|\mathbf{p}_i- \mathbf{p}_j\|}) \right)\! \in\! [0, \frac{\pi}{2}]$. The detailed proof is provided in~\cite{qi2024avoiding}.

In terms of Def.~\ref{def:blocking}, an airplane enters blocking mode when the heading angle is symmetrically aligned with the other airplane's heading direction, thereby continuously preventing each other from bypassing. As shown in Fig.~\ref{fig:solution}, blocking arises when one airplane’s cruising angle falls within the yellow arc and the other’s lies within the red arc. In these situations, the two airplanes tend to choose mirror-image avoidance directions, forcing both airplanes into blocking modes. The formal condition is defined as follows.
\begin{lem}
Consider a two-airplane system $\mathcal{A} = \{A_1, A_2\}$ in which each airplane is equipped with a safety filter in Eq.\eqref{eq:sf}. Both $A_1$ and $A_2$ are in blocking mode at a time instant $T$ $\Iff$ $\exists s \in \{-1, +1\}$, $s\measuredangle(\theta^*_1-\beta_1^2(T)) \in [0, \Delta)$ and $-s\measuredangle(\theta^*_2-\beta_2^1(T)) \in  [0, \Delta)$.
\label{th:block} 
\end{lem} 
\noindent The proof of blocking conditions can be found in~\cite{qi2024avoiding}. We can see that the desired cruising angle $\theta_i^*$ plays a crucial role in the condition of blocking. We propose to adjust $\theta_i^*$ to resolve blocking in the next section.


\subsection{Nonlinear opinion dynamics for blocking resolution}

\begin{figure}[htp]
    \centering
    \includegraphics[width=0.28\textwidth]{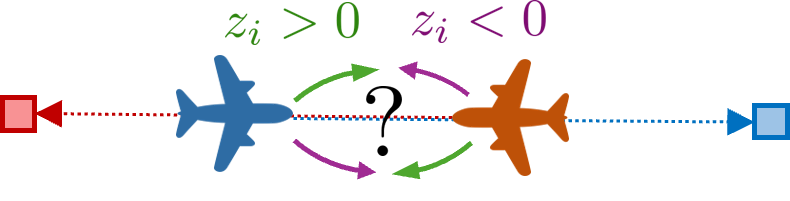}
    \caption{Options in a two-airplane encounter.} 
    \label{fig:opinion}
\end{figure}

Resolving the blocking phenomenon requires the collaboration between both airplanes. In a two-airplane encounter, each airplane can pass the other on either the left or the right, creating four possible maneuver combinations. As Fig.~\ref{fig:solution} illustrates, the conflict can be resolved only when both airplanes choose the same bypass side, producing a clockwise or counter-clockwise swap. For such safety-critical situations, we leverage the nonlinear opinion dynamics to enable two airplanes to quickly reach cooperative decisions on selecting a bypass side, as shown in Fig.~\ref{fig:opinion}.

Let $z_i\!\in\!\mathbb{R}$ denotes the opinion state about its bypass side for $A_i$. The opinion is neutral when $z_i\!=\!0$. A positive opinion $z_i\!>\!0$, implies that the agent $i$ favors right side, while a negative opinion $z_i\!<\!0$ indicates that $A_i$ favors left side. To achieve collaboration, the opinion formulation must incorporate the other airplane's opinion state, denoted by $z_{j}$. Following~\cite{leonard2024fast,cathcart2023proactive}, we define the opinion dynamics as
\begin{equation}
    \dot{z}_i = -d z_i + u_i \tanh \left(\alpha z_i + \gamma z_{j} + b_i\right), 
\label{eq:opinion}
\end{equation}
for $i,j\!\in\!\{1,2\}$ and $i\!\neq\!j$, where $d>0$ is a damping coefficient, $\tanh$ function acts as a nonlinear saturation function to facilitate fast and flexible decision-making, and the attention $u_i\!\geq\!0$ is a tuning variable to be designed, reflecting $A_i$'s real-time attention about blocking situations. The parameters $\alpha\!>\!0$ and $\gamma\!\in\!\mathbb{R}$ serve as weights that determine the influence of $z_i$ and $z_j$ on opinion update of $A_i$, respectively. A positive value $\gamma\!>\!0$ facilitates consensus decision making between the two airplanes, while $\gamma\!<\!0$ leads to dissensus. Therefore, to ensure that both airplanes select the same bypassing side, $\gamma$ should be positive. The external bias parameter $b_i$ represents prior preference. A positive bias $b_i\!>\!0$ represents a preference for right bypassing, while a negative bias favors the left side. 

To tailor the opinion dynamics for blocking resolution, we introduce an attention function that increases as the airplanes approach a blocking mode. Guided by Def.~\ref{def:blocking}, the attention parameter $u$ is designed as the following function
\begin{equation}
u_i = \frac{ k_1 \mathbf{1}\left(g(\mathbf{p}_i, \mathbf{p}_j, \theta_i^*)<0\right)}{\|\dot{\beta}_i^j\| + k_2}, 
\label{eq:attention}
\end{equation} 
\noindent for $i,j\!\in\!\{1,2\}$ and $i\!\neq\!j$, where $\mathbf{1}(\cdot)$ is an indicator function, $k_1\!>\!0$ and $k_2\!>\!0$ are constant gains. \qi{The attention $u_i\!=\!0$ indicates that $A_i$ is in \textit{cruising mode}; $0\!<\!u_i\!<\!\frac{k_1}{k_2}$ in \textit{avoiding mode}; and $u_i\!>\!\frac{k_1}{k_2}$ when in \textit{blocking mode}, for $i\!\in\!\{1,2\}$.}

To break the blocking conditions described in Lemma~\ref{th:block}, the nominal angle in the safety filter should be adjusted by embedding the opinion state $z_i$,
\begin{equation}
    \theta^{n*}_i = \textstyle \theta_i^* + \|\tanh{(k_z z_i)}\| \measuredangle(\beta_i^j-\theta_i^*) + \tanh(k_z z_i) \frac{\pi}{2},
    \label{eq:adjust_angle_2}
\end{equation}
where $i,j\!\in\!\{1,2\}$ and $i\!\neq\!j$ and $k_z$ is a large enough positive gain such that $z_i \!> \!\epsilon\! >\!0$, $\tanh(k_z z_i)\!\approx\! 1$. When $z_i$ is close to $0$, the nominal angle approximates the cruising angle, $\theta^{n*}_i\!\approx\!\theta_i^*$. As $\tanh(z_i)$ converges to $\pm 1$, $\theta^{n*}_i$ converges to $\beta_i+\frac{\pi}{2}$ or $\beta_i-\frac{\pi}{2}$, which reflect a left and right bypass side. By replacing $\theta^{s*}_i \!=\!\mathcal{F}(\mathbf{x}, \theta_i^*)$ with $\theta^{s*}_i \!=\!\mathcal{F}(\mathbf{x}, \theta_i^{n*})$, the blocking is resolved as long as the opinion states $z_1$ and $z_2$ can guide both airplanes to select the same bypass side when approaching the blocking in terms of Lemma~\ref{th:block}. This will be verified subsequently.

\subsection{Blocking-free analysis}~\label{sec:analysis}
This section investigates the blocking-free guarantee in the two-airplane system achieved through nonlinear opinion dynamics~\eqref{eq:opinion}, attention function~\eqref{eq:attention}, and the opinion-guided angle~\eqref{eq:adjust_angle_2}. For analytical simplicity, we consider the case where both agents have no prior preference, i.e., $b_1 \!=\!  b_2 \!=\! 0$, and the gain parameters satisfy $\alpha \!=\! \gamma \!=\! \kappa>0$. Under these assumptions, the two-agent opinion dynamics reduce to:
\begin{equation}\label{eq:NOD}
\left\{ \! \begin{array}{l} \textstyle
\dot{z}_1\!=\! -d z_1\!+\!u_1 \tanh \left( \kappa z_1\!+\!\kappa z_2\!\right) \\
\dot{z}_2\!=\!-d z_2\!+\!u_2 \tanh \left(\kappa z_2\!+\!\kappa z_1\!\right)
\end{array}\right.
\end{equation}
\noindent \qi{The adjacency matrix for this system is given by $A\!=\! \textstyle \begin{bmatrix}0 & 1\\ 1& 0\end{bmatrix}$}. Next, we will demonstrate how this opinion dynamics can enable collaborative decision-making between two options (e.g., left/right bypass turns) for two airplanes, given the positive gains $k_1, k_2, d, \kappa $ satisfying the following condition,
\begin{equation}
\frac{k_1}{k_2}=\frac{d}{2\kappa}+\epsilon
\end{equation}
with $\epsilon$ a small positive scalar.

\begin{figure}[tb]
    \centering
    \includegraphics[width=0.40\textwidth]{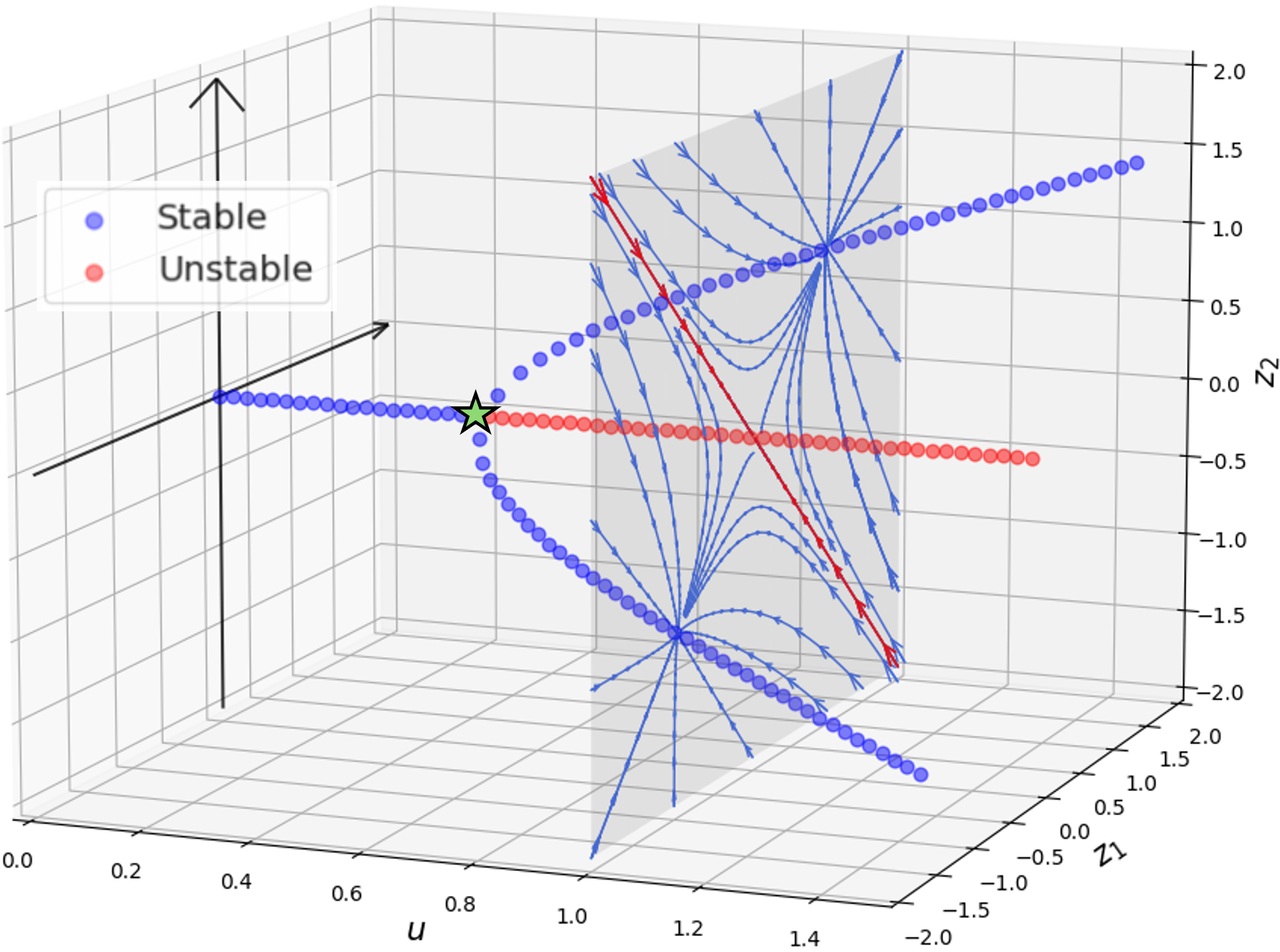}
    \caption{Bifurcation diagram with the slide at $u\!=\!1$ displaying the phase portrait. The pentagram marker denotes the bifurcation point $u^*\!=\!0.5$. The parameters used are $d=\alpha=\gamma=1$, and both agents are unbiased. 
    }
    \label{fig:bifur}
\end{figure}

Following a similar analysis in~\cite[Corollary IV.1.2]{bizyaeva2022nonlinear}, we conclude that, for the system~\eqref{eq:NOD}, the neutral opinion $\boldsymbol z\!=\![z_1\ z_2]^\top\!=\!\mathbf{0}$ is a locally exponentially stable equilibrium when $u_1\!=\!u_2\!<\!u^*$, and becomes an unstable equilibrium when $u_1\!=\!u_2\!>\!u^*$, \qi{where the critical value is given by $u^*\!=\!\frac{d}{2\kappa}$.} The system~\eqref{eq:NOD} undergoes a supercritical pitchfork bifurcation at $u_i\!=\!u^*$, where two branches emerge from $\boldsymbol z\!=\!\mathbf{0}$. The corresponding bifurcation branches are tangent at $\boldsymbol z\!=\! \mathbf{0}$ to the subspace $\text{Span}(\boldsymbol{v}_{max})$, where $\boldsymbol v_{max}\!=\![1 \ 1]^\top$ is an eigenvector associated with the largest eigenvalue of the adjacency matrix, $A$. \tang{To illustrate this, Fig.~\ref{fig:bifur} shows an example of the equilibrium distribution for  the opinion dynamics as the shared attention $u_1=u_2=u$ varies.}

In particular, when the two airplanes are in blocking mode, the attention function~\eqref{eq:attention} satisfies $u_1\!=\!u_2\!=\!\frac{k_1}{k_2}\!>\!u^*$. The inner product $\langle\boldsymbol{z}, \boldsymbol{v}_{max}\rangle \!=\! z_1+z_2$ converges to either the positive branch $z_{1a}^*+z_{2a}^*\!>\!0$ or the negative branch $z_{1b}^*+z_{2b}^*\!<\!0$. Moreover, due to the dynamics~\eqref{eq:NOD}, the bifurcation of $(z_1+z_2)$ will drive individual opinions toward the equilibria with the same sign, either $z_{1a}^*\!>\!0$, $z_{2a}^*\!>\!0$ or $z_{1b}^*\!<\!0$, $z_{2b}^*\!<\!0$. This implies that the two airplanes agree on the same bypass direction, resulting in a coordinated swap, either clockwise or counterclockwise, thereby resolving the blocking situation. When one airplane $A_i$ is in the blocking mode and the other $A_j$ is in the cruising mode, the attentions satisfy $u_i\!>\!u^*$ and $u_j\!=\!0$, meaning that only $A_i$ performs the bypass operation and thereby resolves its blocking. When both airplanes are in cruising or avoiding mode, the attentions satisfy $0\!\leq\!u_i\!<u^*$ for $i\!\in\!\{1,2\}$, so the neutral opinion $z_1\!=\!z_2\!=\!0$ is exponentially stable and $\theta_i^{n*}\!=\!\theta^*_i$ for $i\!\in\!\{1,2\}$, meaning that nonlinear opinion dynamics will not affect the behavior of airplanes. To conclude, the bifurcation of opinion dynamics can be activated as needed to guide the safety filters in resolving blocking phenomena, thereby ensuring both safety and blocking resolution throughout the process.

\subsection{Communication-free implementation}

The above analysis assumes that each airplane knows the other’s opinion state, which is unrealistic when inter-airplane communication is unavailable. Therefore, we design an intention estimator for each airplane to estimate the other’s evolving opinions. Under realistic uncertainty, the local intention regarding which side the other airplane has selected can be inferred from small deviations of real-time relative bearing angle, given by
\begin{equation}
	\hat{z}_{j} = \theta_{j} - \beta_{j}^i,
	\label{eq:estimate}
\end{equation}
\noindent for $i,j\!\in\!\{1,2\}$ and $i\!\neq\!j$. Even if this estimate is imperfect, the airplane with the stronger opinion can compel the other to adopt its chosen side. Because the bifurcation is inherently ultrasensitive, this opinion-based ``negotiation" yields collaborative decisions between airplanes rapidly. The following simulations validate this behavior.

\section{Simulation Validation}\label{sec:exper}

In the experiments, each airplane is assigned a target destination, denoted by $\mathbf{p}_i^g$ for $A_i\!\in\!\mathcal{A}$. The desired cruising angle is computed as follows: $\textstyle \begin{bsmallmatrix} \text{cos}(\theta_i^*)\\ \text{sin}(\theta_i^*)
\end{bsmallmatrix} = \frac{\mathbf{p}^g_i - \mathbf{p}_i}{\|\mathbf{p}^g_i - \mathbf{p}_i\|}$.  Airplanes can accurately observe the state of the other airplanes and share identical parameters: $v=1$, $d\!=\!3$, $\alpha\!=\!1$, $\gamma\!=\!4$, $k_1=2$, and $k_2=0.1$. Moreover, since we assume airplanes do not have predefined rules, airplanes are unbiased by default, i.e., $b_i\!=\!0$ for $A_i\!\in\!\mathcal{A}$. To model environmental noise, we add a Gaussian noise term $\omega_\theta$ to the heading angle $\theta_i$. To validate the ultrasensitivity on the bifurcation point, the noise is set to a relatively small level, $\omega_\theta\!\sim\!\mathcal{N}(0, 0.01)$. The related simulation videos are accessible at the provided link~\footnote{{\small\url{https://youtu.be/7TjrVCJQswM}}}.

\begin{figure}[t]
    \centering
    \subfigure[Clockwise swap]{
        \includegraphics[trim=100 0 100 0, clip, width=0.465\linewidth]{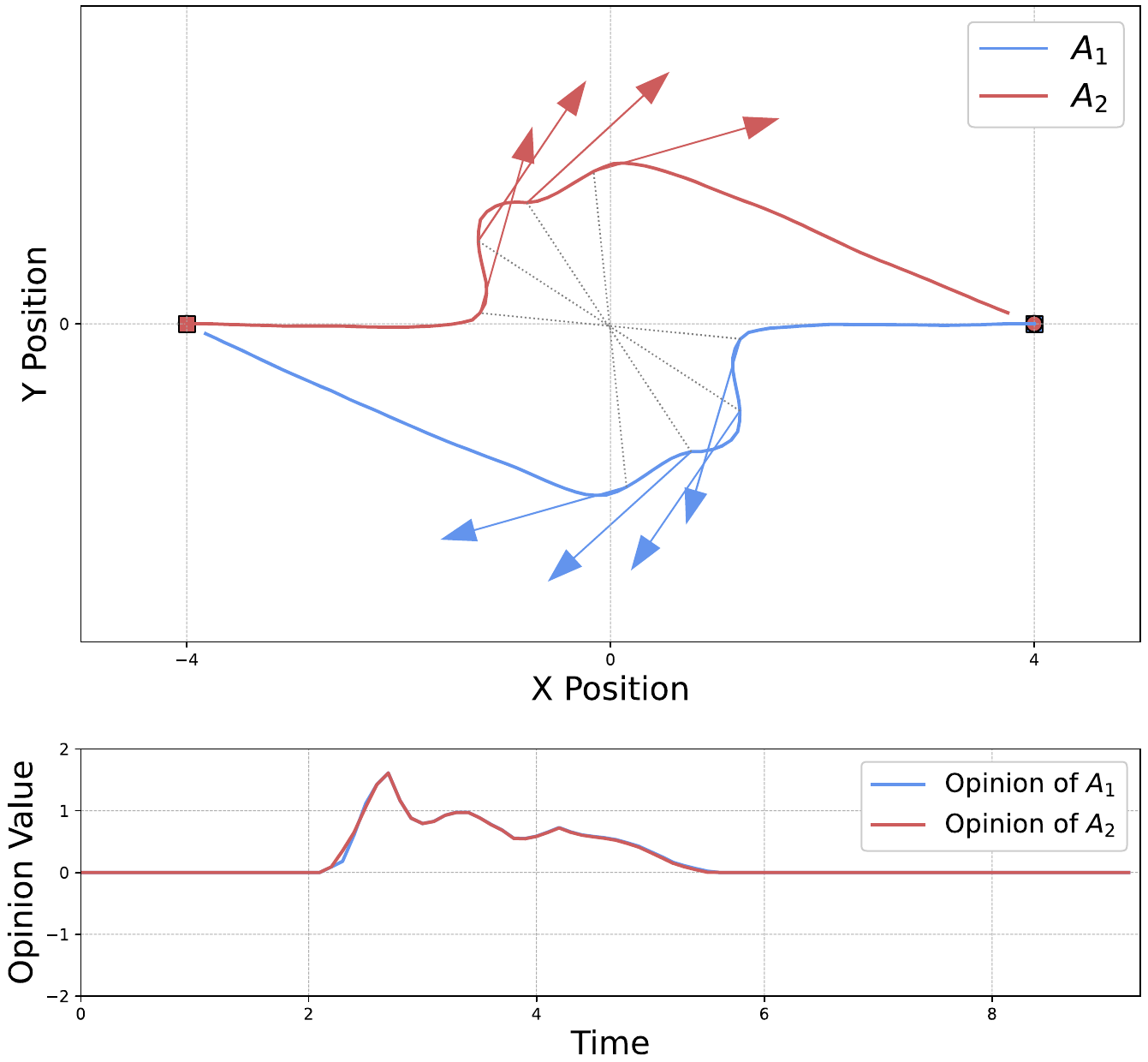}
    }
    \subfigure[Anti-clockwise swap]{
        \includegraphics[trim=100 0 100 0, clip, width=0.465\linewidth]{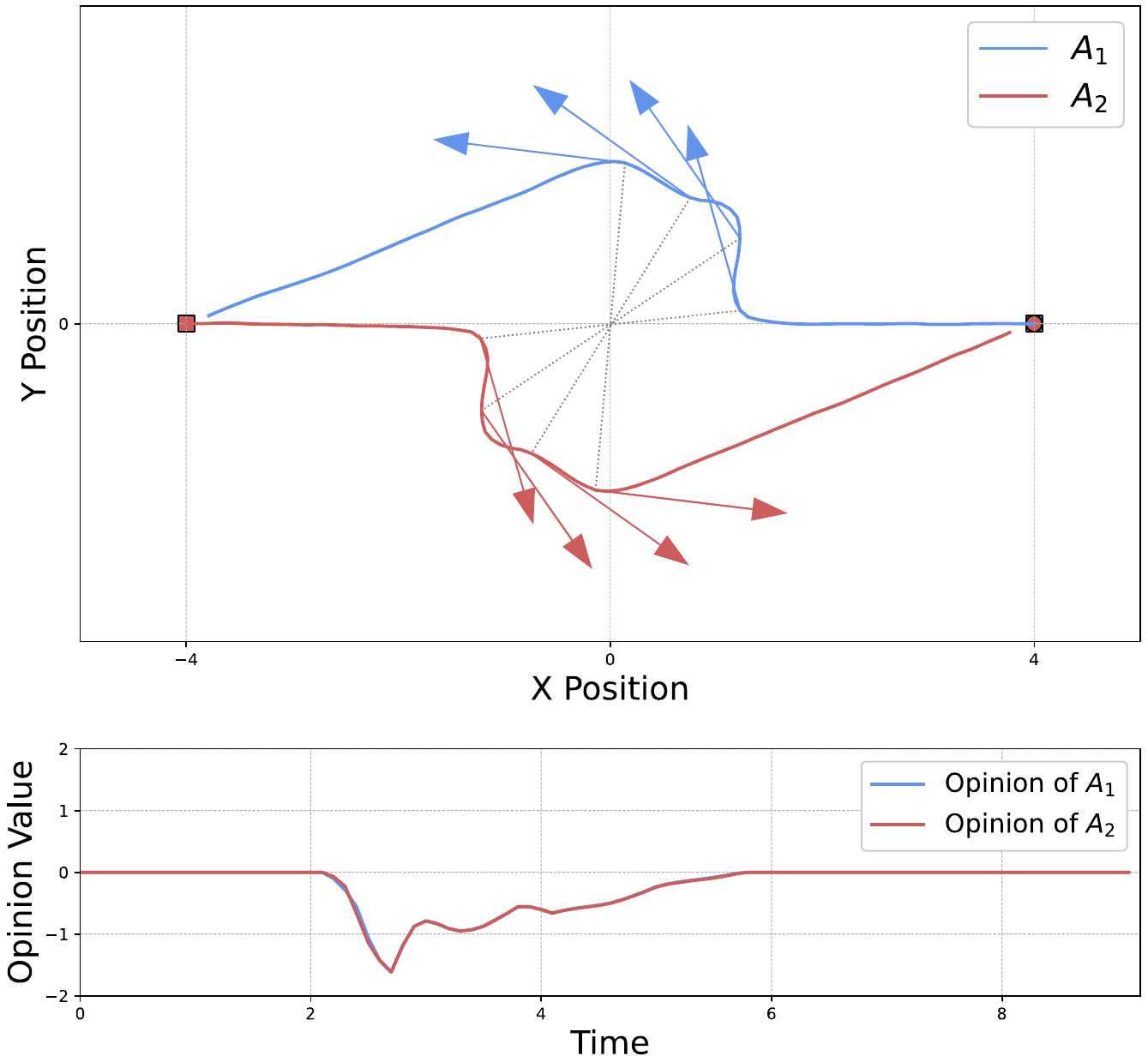}
    }
    \caption{Two resolutions behavior randomly generated in two-airplane encounters. The upper panels show the airplanes' trajectories, while the bottom panels depict the evolution of opinion states. The arrows indicate the nominal heading $\theta^{n*}_i$ adjusted by the opinions, while the gray dashed lines connect the positions of two airplanes during the activation of opinions. Circle and square markers denote initial and target positions, respectively. }
    \label{fig:TE_sim_1}
\end{figure}

We first test the effectiveness of opinion dynamics~\eqref{eq:opinion} for a two-airplane system in a completely symmetric scenario, as shown in Fig.~\ref{fig:TE_sim_1}. Without the opinion-guided adjustment, such a symmetric situation results in a catastrophic deadlock since both airplanes consistently and simultaneously select the opposite bypassing side, as identified in~\cite{qi2024avoiding}. The cooperative behavior of the proposed method shown in Fig.~\ref{fig:TE_sim_1} demonstrates that the opinion dynamics effectively break the symmetry and enable collaborative decision-making in a decentralized manner. Moreover, the choice between clockwise and anticlockwise swaps is randomly determined by the online noise. Such rapid consensus decision making demonstrates the bifurcation property discussed in Sec.~\ref{sec:analysis}. 


\begin{figure}[ht] \vspace{-3mm}
    \subfigure[$b_1=b_2=0$]{\includegraphics[trim=0 25 0 20, clip, width=0.48\linewidth]{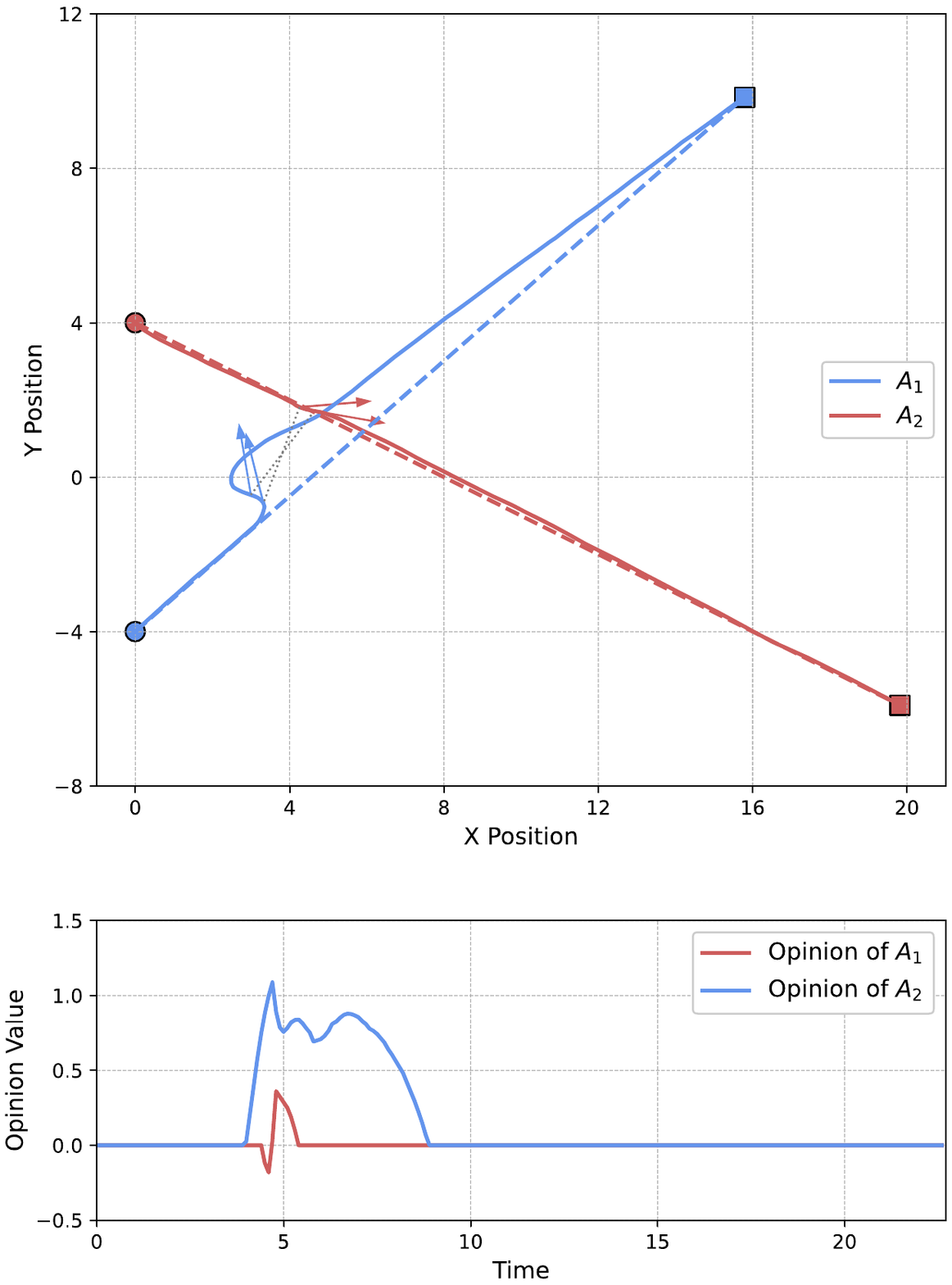}\label{fig:zero_bias}
    }
    \subfigure[$b_1=-10, \, b_2=0$]{\includegraphics[trim=0 10 0 0, clip,width=0.47\linewidth]{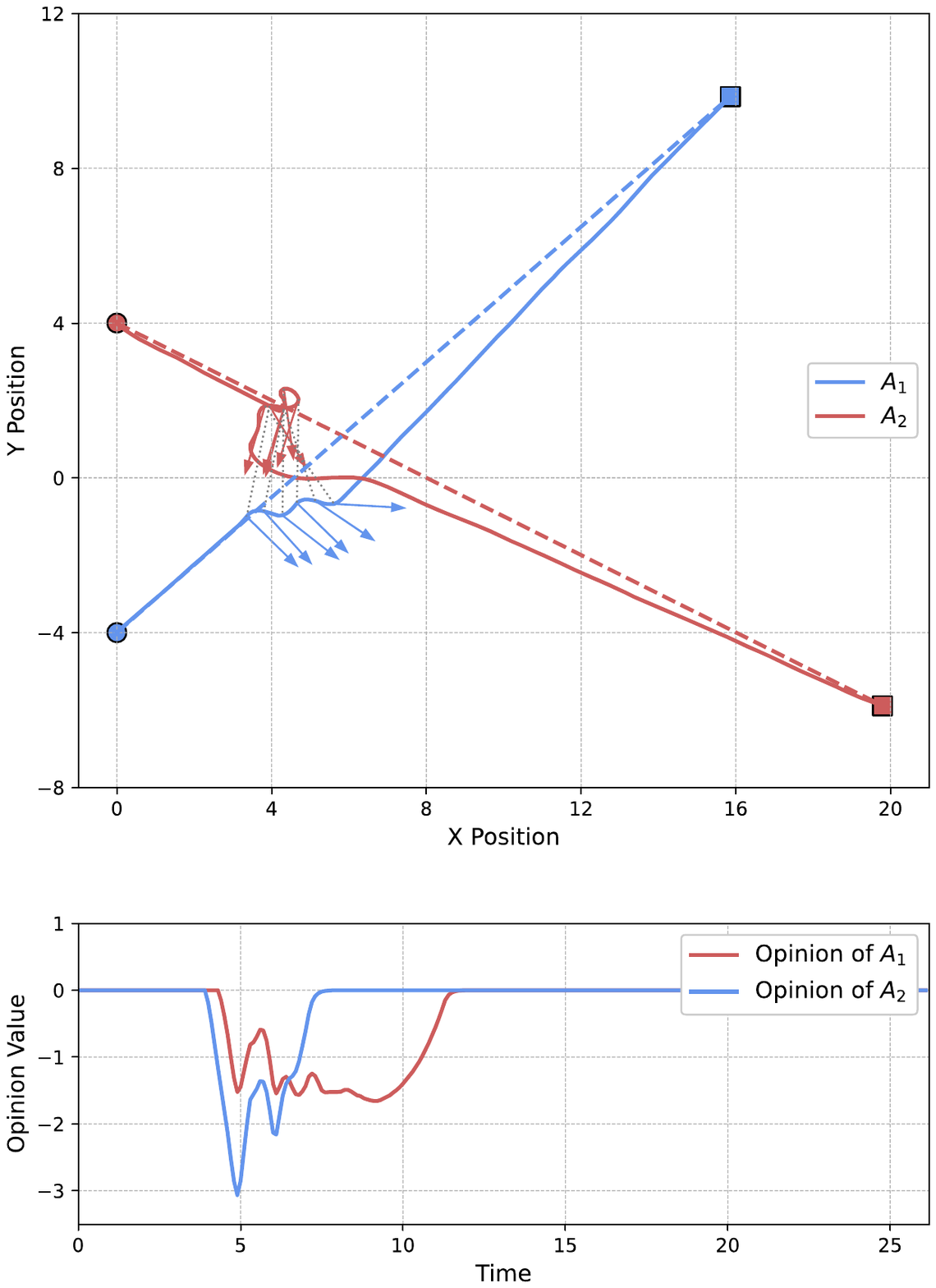}\label{fig:bias}
    }\vspace{-2mm}
    \caption{Resolution for the blocking phenomenon in Fig.~\ref{fig:blocking_sim}. 
    }
    \label{fig:TE_sim_2} \vspace{-2mm}
\end{figure}

The resolution behavior for the blocking phenomenon in Fig.~\ref{fig:blocking_sim} is depicted in Fig.~\ref{fig:TE_sim_2}. Unlike the symmetric case in Fig.~\ref{fig:TE_sim_1}, where both airplanes randomly choose either side for bypassing, airplane $A_1$ consistently chooses to take a selfless detour, while $A_2$ deviates slightly from its reference path, as shown in Fig.~\ref{fig:zero_bias}. This occurs because $A_2$ approaches the intersection point earlier than $A_1$, thereby giving it the advantage to maintain its intended path. The resulting swap maneuver proves to be the most efficient. When an external bias is set to force $A_1$ to bypass from the right side, as shown in Fig.~\ref{fig:bias}, $A_2$ adapts its decisions in time. These results show that the opinion-based approach facilitates rational and adaptive negotiation between agents, which is generally absent from rule-based methods.

\begin{figure}[t]
    \centering
    \includegraphics[width=0.90\linewidth]{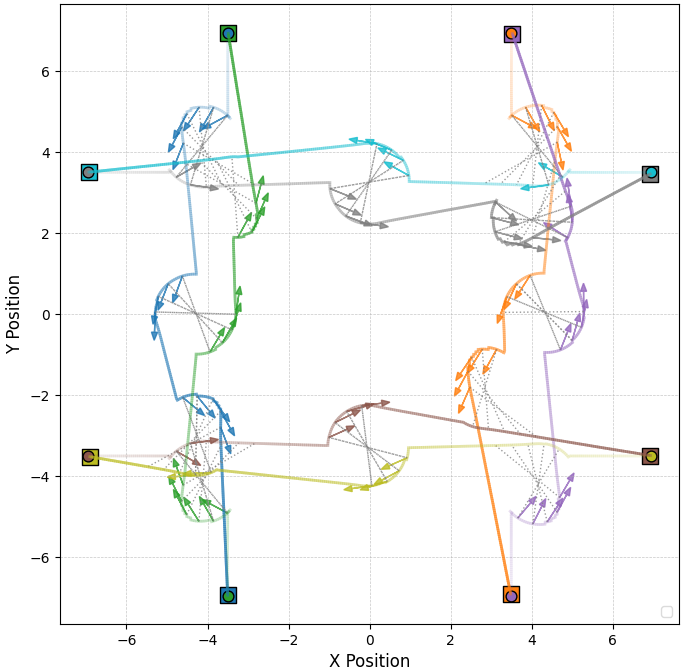}
    \caption{Large-scale simulations involving 8 airplanes, with each airplane encountering three others during flight. The used markers are identical to those in Fig.~\ref{fig:TE_sim_1}. 
    }
    \label{fig:LS_sim}
\end{figure}

Using the opinion-guided approach, the flying time of $A_2$ in Fig.~\ref{fig:zero_bias} is reduced by $22.3\%$ compared to the blocking path in Fig.~\ref{fig:blocking_sim}. Although $A_1$ actively adopts a selfless detour, it also saves an additional $5\%$ in flying time. This time-saving benefit is not limited to this case study. We evaluated the proposed approach in 200 randomly generated encounter scenarios. In these tests, the initial points of airplanes were fixed, while their target positions were randomly selected within regions prone to blocking.  The results indicate that the blocking strategy saves an average of $19.7\%$ flying time for each airplane. Importantly, no airplane entered blocking mode, and all safety constraints were maintained throughout the simulations. Thus, both the safety guarantee and the performance benefits of the opinion-guided blocking resolution are verified. In addition, we also evaluate the proposed method in large-scale traffic scenarios. In Fig.~\ref{fig:LS_sim}, eight airplanes are assigned tasks of switching their positions, with each airplane sequentially encountering three others before reaching its target. The smooth and safe trajectories indicate that all the encounters are properly managed. Note that each encounter involves only two airplanes at a time.

\section{Conclusion}
This paper tailors opinion dynamics for airplanes to resolve blocking phenomena without relying on communication or fixed rules, and integrates it into a safety control framework. Our approach ensures safety and blocking-free resolution, as validated through extensive simulations. The proposed integration of the bifurcation property of nonlinear opinion dynamics with the safety guarantee of a CBF-based controller offers practical insights for advancing DAA systems in real airplanes. 

This paper provides a bifurcation analysis of the designed opinion dynamics. Future work will involve a rigorous analysis of the integrated system comprising the opinion dynamics and the intention estimator. Furthermore, the proposed method and its guarantees are limited to two-airplane encounters modeled with simplified dynamics. Future research will explore learning-based opinion networks to enable more sophisticated cooperation in challenging multi-airplane encounters with realistic dynamics models. 


\bibliographystyle{ieeetr}        
\bibliography{MyBib}    
\balance

\end{document}